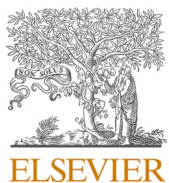
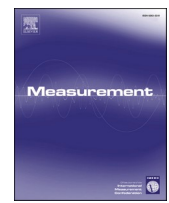

# Portable, cost-effective UV–vis-NIR microspectrophotometer for absorption and fluorescence microscopy and spectroscopy

Negar Karpourazar, Keyvan Khosh Abady, Peter M. Rentzepis *

*Department of Electrical and Computer Engineering, Texas A & M University, College Station, TX 77843, USA*

ABSTRACT

This article describes the design and construction of a portable, compact, and cost-effective microspectrophotometer (MSP) that operates in the range of (200-800 nm). This microscope-spectrophotometer records high-resolution absorption and emission spectra in situ. The dual-head design of this MSP enables simultaneous real-time imaging and spectral recording of heterogeneous samples with high selectivity and micrometer spatial resolution. Our compact, portable MSP design reduces construction costs by more than 20 times compared to commercial benchtop alternatives, primarily due to its innovative illumination system and microscope objective design. The performance of the UV–vis-NIR MSP was confirmed by comparing the absorption and fluorescence spectra of an aqueous solution of $Ru(bpy)_3^{2+}$ obtained with our system to those measured by commercial spectroscopic systems. The high accuracy and reliability of our system in measuring absorbance and fluorescence were confirmed by R-squared values of 0.998 and 0.990, respectively, from colorimetric and fluorometric tests. The MSP was further used to record absorption and fluorescence spectra from a variety of samples, including dyes and protein crystals, in both the solution and solid state, as well as individual living cells. This compact instrument is ideal for rapid, in situ spectroscopic measurements and is expected to find on-site applications across various fields, such as environmental monitoring, biological research, forensic analysis, and materials characterization.

## 1. Introduction

Spectroscopy is a reliable, fast, and non-destructive analytical technique that utilizes the interaction between matter and electromagnetic radiation to detect and characterize the chemical composition of various materials [1–6]. The microspectrophotometer (MSP), which integrates microscopy with spectroscopy, can be configured to record various types of spectra, including absorption, fluorescence, Raman, infrared, and nuclear magnetic resonance [7–12]. MSPs provide several advantages over conventional spectroscopic systems. Notably, these microscope-based systems enable the recording of spectra with micron-scale spatial resolution, while simultaneously imaging microscopic features and spatial heterogeneity of the sample [13–15]. This capability is crucial for in situ spectroscopic analysis of heterogeneous samples, as it enables the selective recording of spectra from individual components without interference from other regions, thereby improving analytical accuracy [16]. Additionally, MSPs are capable of recording spectra from small sample volumes with high signal-to-noise ratios, whether in liquid or solid form. Furthermore, MSPs require little to no sample preparation. These unique capabilities enable a wide range of applications in various scientific fields. In biological research, MSPs have been used to record the absorption spectra of individual living cells and their organelles [17–19]. In forensic science, MSPs are used to analyze microscopic colored materials or substances with UV absorbers, such as dyes, inks, paints, and fibers [20–22]. Furthermore, MSPs have important applications in environmental monitoring, medical diagnostics, material characterization, microfluidics analysis, biotechnology research, and quality control in the food and pharmaceutical industries [23–28].

Despite their many capabilities, MSPs have several major limitations, including high cost, large size, intricate design, and complex operational procedure [29,30]. These limitations hinder their on-site application, requiring samples to be transported to dedicated laboratories for analysis. This expensive and time-consuming process, from sample collection to analysis, presents significant challenges, particularly in healthcare, environmental monitoring, and forensic investigations, where rapid detection is crucial [31,32]. Additionally, most MSPs operate within the narrow spectral range of visible light, which further limits their potential applications. This limitation is also significant because the energies associated with the electronic and vibrational spectra of most compounds and crystal structures fall within a broad spectral region, ranging from ultraviolet (UV) to near-infrared (NIR) [33]. For instance, Yu et al. (2023) developed a handheld spectrophotometer for colorimetric and fluorescence analysis [34], while Fang et al. (2025) introduced a portable device for similar applications using LAMP technology [35]. Although these devices are portable, low-cost, and user-friendly, they lack microscopy capabilities and are unable to selectively record spectra from specific micron-sized regions of a sample. Moreover, their spectral range is limited to the visible region (405-690 nm).

To address these limitations, this article describes the design and construction of a novel, compact, portable and cost-effective microspectrophotometer that operates in the broad spectral range, from UV to NIR (200-800 nm). The compact design of our MSP reduces total construction costs by more than 20-fold compared to commercially available benchtop alternatives with similar capabilities. Even without considering the expenses of the spectrometer and camera, which can be upgraded independently to higher-end components for improved resolution, the innovative design of the illumination system and microscope objective results in a substantial cost reduction. The development of this UV–vis-NIR MSP was based on our previously designed compact and economical UV microscope [36]. The presented dual-head MSP provides enhanced capabilities for recording absorption spectra by utilizing a compact illumination source that emits a continuous spectrum ranging from UV to NIR (200-2200 nm). In addition, fluorescence spectra can be recorded using the same system by integrating a custom-built LED array into the illumination path and an appropriate color glass filter into the detection path, without the need for additional components or design modifications. The dual-head setup of this device enables integration of a camera and a spectrometer into the detection system. This configuration enables real-time imaging of the sample while simultaneously recording spectra with high selectivity.

To validate the performance and applicability of our UV–vis-NIR MSP, spectra were recorded from various well-studied materials. These materials were chosen for their wide range of applications and the availability of comparable research data, which allowed for a thorough evaluation of our MSP's accuracy and performance. The tested materials included $Ru(bpy)_3^{2+}$ dyes and bacteriorhodopsin proteins, both in solution and in solid state, as well as living cells of Chlamydomonas reinhardtii. $Ru(bpy)_3^{2+}$ is an organometallic dye known for its photochemical stability, redox properties, and ability to exhibit both electroluminescence and chemiluminescence, making it useful for various applications such as photocatalysis and biosensing [37–39]. Bacteriorhodopsin is a membrane protein that acts as a light-driven proton pump, making it an ideal model system for the study of protonation reactions [40,41]. The crystalline structure of bacteriorhodopsin exhibits exceptional stability against chemical and thermal degradation [42]. In addition, bacteriorhodopsin possesses unique optical properties, including a large absorption cross-section, high quantum efficiency, and photochromic characteristics, which make it ideal for applications such as optical memory devices, solar energy harvesting, and sensor technologies [43–48]. Chlamydomonas reinhardtii (C. reinhardtii) is a photosynthetic unicellular biflagellate microalga that serves as a model organism for studying important biological and biochemical processes in plants and eukaryotes [49,50]. Research using this organism focuses on diverse topics such as photosynthesis, respiration, metabolism, biosynthetic pathways, hydrogen production, and chloroplast gene expression [51–55]. The results from these diverse samples demonstrate that our MSP accurately records both absorption and fluorescence spectra of various liquids, solids, and living biological samples.

The portable, compact, reliable, and economical UV–vis-NIR microspectrophotometer, presented in this paper, is particularly advantageous for applications requiring accurate, user-friendly, and cost-effective spectroscopic systems for the in situ analysis of heterogeneous samples. This includes rapid and precise in-field experiments in areas such as pathogen detection (including bacteria and fungi), forensic analysis, semiconductor industry, environmental pollution monitoring, cancer research, DNA analysis, protein crystal studies, microfluidics analysis, and food and pharmaceutical industries.

## 2. Materials and methods

### 2.1. Light sources

A novel combination of light sources was employed in our MSP to provide the required illumination for both absorption and fluorescence spectroscopy in a compact system. For absorption spectroscopy, broad continuous illumination was provided by a dual-lamp system. This system comprised a compact, high-intensity deuterium ($D_2$) lamp with a see-through design (Hamamatsu, Model L10804,), emitting in the range of 200 nm to 400 nm, coupled to a tungsten-halogen lamp covering the range of 400 nm to 2200 nm. The see-through design of the $D_2$ lamp enabled the simultaneous access to UV, visible, and NIR spectra, while enhancing the system's compactness and ease of operation. For fluorescence excitation, a novel, compact LED array system with customizable illumination wavelengths was designed, constructed and integrated into the dual-lamp system. Our LED array setup incorporates high-power 3 W LEDs emitting at wavelengths of 365 nm, 395 nm, 420 nm, and 520 nm. These LEDs can be easily replaced with ones at the desired wavelengths.

Alternatively, arc discharge lamps, such as mercury and xenon lamps, are viable options for MSP light sources. These lamps provide a broad spectrum (200-600 nm) and offer a high radiance output. However, mercury lamps have distinct high-intensity emission lines in their spectral output, making them unsuitable for absorption spectroscopy, where a uniform continuous spectrum is required to ensure accurate and reliable measurements. The high-intensity emission lines of a mercury lamp are commonly used as excitation sources in fluorescence spectroscopy. However, this approach limits the excitation wavelengths to the specific mercury emission lines, resulting in significant spectral gaps, especially between 436 nm and 546 nm, and beyond the 579 nm region. Although xenon arc lamps provide the continuous and uniform UV-visible spectrum needed for absorption measurements, neither mercury nor xenon lamps are suitable for constructing a compact microspectrophotometer because of their large size and the requirement for specialized lamphouses and power supplies. Additionally, they may present a potential risk of explosion and higher costs.

### 2.2. Optical components

Simple, low-cost fused silica hemispherical lenses were used as microscope objectives, which efficiently transmit light between 200 nm and 2 μm with minimal attenuation. In addition, the high numerical aperture of these lenses enables the formation of high-resolution images. In our experiments, a fused silica hemispherical lens with a diameter of 4 mm (Edmund optics #67–396) was used as the microscope objective. Higher magnification can be achieved by using a ball lens with a smaller diameter.

To incorporate a dual-head microscope, an internal beam splitter that transmits in deep UV is necessary within the microscope head. Conventional beam-splitters used in dual-head light microscopes are typically made of glass, which does not transmit UV light. Custom-made UV-transmissive alternatives, however, often come at prohibitively high costs. In our instrument, a UV fused silica plate beam splitter (Thorlabs #BSW20) was utilized. This beam splitter provides a consistent 50:50 beam splitting ratio within the 250-450 nm spectrum while exhibiting a higher transmission-to-reflection ratio at wavelengths outside this range. An alternative option is a polka dot beam splitter with a $CaF_2$ substrate that provides a 50:50 beam splitting ratio from 180 nm to 8.0 μm. In our MSP, biconvex fused silica lenses with focal lengths of 50 mm (Thorlabs #LB4096) and 35 mm (Thorlabs #LB4879) were utilized as the tube lens and condenser lens, respectively. For fluorescence measurements, a long-pass colored glass filter with a cut-on wavelength of 550 nm (Thorlabs #FGL550) was employed.

### 2.3. Camera

A CMOS digital eyepiece microscope camera (AmScope MD35), sensitive to visible light, was used as the primary imaging system for our MSP. In addition, a conventional CMOS image sensor, typically utilized in webcams, was modified to achieve deep ultraviolet sensitivity. This process involved removing the UV and IR blocking filters and debayering the sensor by eliminating the color filter arrays from the sensor's surface, thereby enhancing its spectral response. The Bayer filter can be removed by mechanical stripping or dissolving it with appropriate chemical agents while maintaining the integrity of the sensor. The resulting UV-sensitive camera significantly improves spatial resolution and contrast, making it highly effective for imaging bacteria, RNA/DNA, and protein samples, which demonstrate strong absorption in the UV region [33,36].

### 2.4. Detector

Previous studies focusing on the development of microspectrophotometers have primarily relied on photomultiplier tubes (PMTs) as photodetectors. However, PMTs in MSP systems require the use of a scanning monochromator. Although compact PMTs are commercially available and reflective holographic gratings have enabled compact, precise, and broad-spectrum monochromators, the need for mechanical drive systems and associated power supplies results in physically large systems with slow acquisition rates. In contrast, utilizing a miniature spectrometer eliminates the need for monochromators, reduces the system size, and shortens acquisition times to fractions of a second. Our MSP system incorporates a USB2000 + miniature fiber optic spectrometer equipped with a Charge-Coupled Device (CCD) array detector. A step-index multimode optical fiber with a 200 μm core diameter and a numerical aperture (NA) of 0.22 (Thorlabs #FG200UEA) was used to couple light into the spectrometer. This configuration allows for scanning across the spectral range from 200 nm to 800 nm. The measurable range of the MSP is determined by the choice of light sources and optical components, which in our design span wavelengths from 200 nm to 2 μm. However, for the experiments conducted, the spectrometer's detection limit constrained the operational range to 800 nm. This detection range can be extended to longer wavelengths by replacing the current spectrometer with one that meets specific experimental needs.

### 2.5. Spatial resolution assessment

The spatial resolution of the MSP system was determined using a 2″ × 2″ fused silica USAF 1951 resolution target (Edmund Optics #57896). The smallest resolvable group and element on the target were identified, and the corresponding resolution was calculated using the following equation:

$$\text{Resolution (lp/mm)} = 2^{\left(\text{Group number} + \frac{\text{Element number} - 1}{6}\right)} \quad (1)$$

This equation yields the number of line pairs (black and white) that can be resolved per millimeter. The width of each line pair in micrometers was then calculated using the equation:

$$\frac{1000}{\text{Resolution (lp/mm)}} \quad (2)$$

The value obtained from Eq. (2) represents the minimum optical resolution achievable by the system.

To measure the field of view (FOV), a line on the USAF 1951 target corresponding to a width of 100 pixels on the camera screen was selected. Since the width of a single line is half of the value obtained from Eq. (2), the FOV in micrometers was calculated by counting the number of pixels spanning the full horizontal and vertical ranges of the image and then converting this value to micrometers. Using this technique, we measured the FOV with both standard 40× and 4 mm half-ball objective lenses. The magnification of the 4 mm half-ball objective lens was then determined by scaling the known 40× magnification using the ratio of the measured FOVs.

In the MSP system, light collected by the objective lens is split between the imaging camera and the spectrometer. While the camera defines the overall FOV, only a small portion of this area is directed to the spectrometer via a fiber optic cable. The size of this spectrally sampled region is influenced by several factors, including differences in optical path length between the camera and spectrometer, the numerical aperture of the optical fiber, and the alignment of the fiber within the eyepiece tube. This region ultimately defines the spatial resolution of spectral measurements.

To identify the specific region of the FOV contributing to the spectral signal, a grid overlay of camera was used. A microscope slide containing a black square, sized to match one grid square (100 pixels), was translated across the FOV while spectra were recorded. A significant decrease in spectral intensity was observed when the black square entered the area sampled by the fiber, enabling the precise mapping of the spectrally active region.

To determine the size of this region, the USAF 1951 resolution target was used again, and a line feature with a width matching the sampled area was identified.

### 2.6. Irradiance measurement of light sources

Irradiance measurements were performed to assess the potential photodamage and other light-induced effects caused by each excitation source on sensitive samples. To that effect, the irradiance of the $D_2$ lamp, halogen-tungsten lamp, and four LEDs (365 nm, 395 nm, 420 nm, and 520 nm) was measured at the sample plane under standard operating conditions. A calibrated silicon photodiode (Thorlabs #DET10A2) with a 0.8 mm² active area and a spectral responsivity range of 200-1100 nm was used to measure light intensity. The photodiode was precisely aligned with the sample position to ensure accurate detection. Its photocurrent output was converted to voltage using a transimpedance amplifier. To ensure measurement accuracy and confirm operation within the photodiode's linear response range, at least five different feedback resistor values were tested, and the resulting voltage outputs were confirmed to vary linearly with the feedback resistance values under constant incident power for each light source.

### 2.7. Power supplies

To further enhance portability, our MSP design incorporates a rechargeable lithium-ion battery (Talentcell PB240B1) which provides multiple voltage outputs: 24 V DC for the $D_2$ lamp power supply, 12 V DC for the tungsten-halogen lamp, and 5 V USB for the stepper motor driver module.

### 2.8. Spectrophotometric and colorimetric tests

To validate the spectrophotometric performance of our UV–vis-NIR MSP, absorption spectra of 0.1 mM $Ru(bpy)_3^{2+}$ aqueous solutions were recorded using both the MSP and a commercial Shimadzu UV-1601 spectrophotometer for comparison.

Colorimetry is an accurate, simple and rapid optical detection method for determining the concentration of target substances in solution by measuring and comparing the absorbance of colored compounds [56–58]. To evaluate the precision of our MSP in measuring absorbance and to determine its linear dynamic range (LDR), colorimetric tests were conducted using our device using an aqueous solution of $Ru(bpy)_3^{2+}$ with concentrations ranging from 10 μM to 1 mM.

To ensure consistency and facilitate data comparison, all these measurements were performed using a quartz cuvette with an optical path length of 1 mm. Each measurement was repeated five times.

*2.9. Spectrofluorometric and fluorometric tests*

To evaluate the spectrofluorometric performance of our UV–vis-NIR MSP, fluorescence spectra of 0.1 mM aqueous solutions of $Ru(bpy)_3^{2+}$ were recorded at an excitation wavelength of 420 nm using both the MSP and a commercial Shimadzu RF-6000 spectrofluorometer for comparison.

Fluorometry is a sensitive optical detection method that measures changes in fluorescence intensity, caused by varying the analyte concentrations or chromogenic reactions [59]. To demonstrate the precision of MSP in measuring fluorescence intensity and to determine its LDR, fluorometric tests were performed on $Ru(bpy)_3^{2+}$ aqueous solutions with concentrations ranging from 10 μM to 1 mM.

Consistency across all fluorescence experiments was ensured by using a quartz cuvette with a 1 mm optical path length. Each measurement was repeated five times.

Fluorescence measurements using the MSP were conducted by exciting the samples with a high-power LED. An emission wavelength of 420 nm was used for $Ru(bpy)_3^{2+}$ in both solution and solid-state, while a wavelength of 365 nm was used for C. reinhardtii cells. To ensure accurate fluorescence detection, a long-pass colored glass filter with a cut-on wavelength of 550 nm was placed in the detection path to block the excitation wavelength of the LED while allowing the fluorescence signal to pass through.

*2.10. Absorption and fluorescence analysis of solid samples*

Spectroscopic analysis of solid materials is essential because it provides valuable insights into their intrinsic optical properties and helps to understand how crystal structures and solvents influence absorption bands, offering a comprehensive understanding of material characteristics [60]. Conventional methods for recording solid-state spectra involve embedding the material under study in matrices such as KBr, polymers, gels, or dried and frozen samples [61–63]. However, these techniques have limitations. For instance, KBr pellets are sensitive to moisture and fragile. Their preparation is time-consuming, requires specialized equipment, and often results in inhomogeneous samples, which may lead to baseline drift [64]. Additionally, the conditions of drying, freezing, and sample uniformity can affect the results. Unlike other methods, our MSP directly records spectra from solids without the need for sample preparation. In this study, we demonstrated this approach using crystals of $Ru(bpy)_3^{2+}$ and bacteriorhodopsin. This method not only reduced time-consuming procedures and eliminated the need for specialized equipment but also avoided the introduction of potential artifacts, such as those resulting from sample-matrix interactions, nonuniformity, and solidification conditions.

*2.11. Materials*

The powder of tris (2,2′-bipyridyl) dichlororuthenium (II) hexahydrate ($Ru(bpy)_3^{2+}$) were purchased from Sigma-Aldrich. An aqueous solution of $Ru(bpy)_3^{2+}$ of a 1 mM concentration was prepared using distilled water. This stock solution was then diluted with distilled water to obtain the desired concentrations for colorimetric and fluorometric tests. $Ru(bpy)_3^{2+}$ crystals were placed directly on a quartz plate for solid-state analysis without further sample preparation. Dried samples of $Ru(bpy)_3^{2+}$ were prepared by drying 100 μl of stock solution on a quartz plate at room temperature. The lyophilized bacteriorhodopsin powder with the native sequence of Halobacterium salinarum was purchased from Sigma-Aldrich. The bacteriorhodopsin solution was prepared by dissolving the flakes in a detergent buffer solution consisting of 20 % w/v glycerol, 2 mM $MgCl_2$, 140 mM NaCl, 50 mM HEPES and 1 mM DTT. Approximately 5 ml of buffer was added to 50 mg of DMM detergent. For solid-state analysis, bacteriorhodopsin crystals were placed directly on a quartz plate without additional sample preparation. For the spectroscopic analysis of solid samples, thin crystals were selected. The C. reinhardtii cell culture was purchased from Carolina Biological Supply. C. reinhardtii cells were placed on a quartz plate and covered with a quartz cover slip for analysis.

## 3. Results and discussion

*3.1. Instrument construction*

The body of a basic dual-head microscope was utilized for the construction of the UV–vis-NIR MSP. A photograph and a detailed schematic of the assembled compact UV–vis-NIR MSP are presented in Fig. 1(a) and Fig. 1(b), respectively. The original lighting system of the microscope was replaced by a compact illumination system that integrated a dual-lamp setup and an LED array. For absorption spectroscopy, the dual-lamp illumination system was composed of a see-through deuterium lamp coupled with a tungsten-halogen lamp, providing a continuous spectrum that extended from the UV to the NIR region, as shown in Fig. 1(c). For fluorescence spectroscopy, a novel and compact LED array was developed. The array consists of high-power LEDs with varying wavelengths, mounted on a circular aluminum plate. This plate was positioned between the $D_2$ and tungsten-halogen lamps and featured a small opening that allowed the fixed-position tungsten-halogen light beneath it to pass through for continuous illumination, as shown in Fig. 1(d). The aluminum plate was connected to a stepper motor using a slip ring to prevent wire entanglement. The stepper motor driver module and the LEDs were connected to an Arduino microcontroller controlled by a computer. Arduino was programmed to control the rotation of the plate, align the selected LED or the opening of the halogen lamp with the see-through path of the $D_2$ lamp, and to switch the LEDs on and off. To enhance the portability of the MSP, the illumination system was powered by a rechargeable battery. The $D_2$ lamp power supply was connected to the 24 V DC output of the battery through a buck converter to regulate voltage fluctuations. The power for the tungsten-halogen lamp and the stepper motor driver module was also supplied by the battery, as previously described. The glass condenser lens beneath the microscope stage was replaced with a fused silica biconvex lens to concentrate the UV–vis-NIR irradiation into a cone-shaped beam, providing even and bright illumination across the entire field of view. The microscope objective was constructed using a 4 mm diameter fused silica hemispherical lens. This lens was embedded into a plate and mounted onto a conventional microscope objective housing, from which the internal optics had been removed. This half-ball objective lens was inserted into the microscope turret. The original glass prism beam splitter inside the microscope head was replaced with a UV fused silica plate beam splitter, which was mounted on a custom holder. A biconvex fused silica lens was incorporated as a tube lens within the microscope head to expand the field of view. A CMOS camera was connected to one eyepiece tube of the microscope, while a spectrometer was connected to the other eyepiece tube using optical fibers attached to a holder. Table 1 provides a summary of the components used in the MSP system, including their functions and operating wavelength ranges.

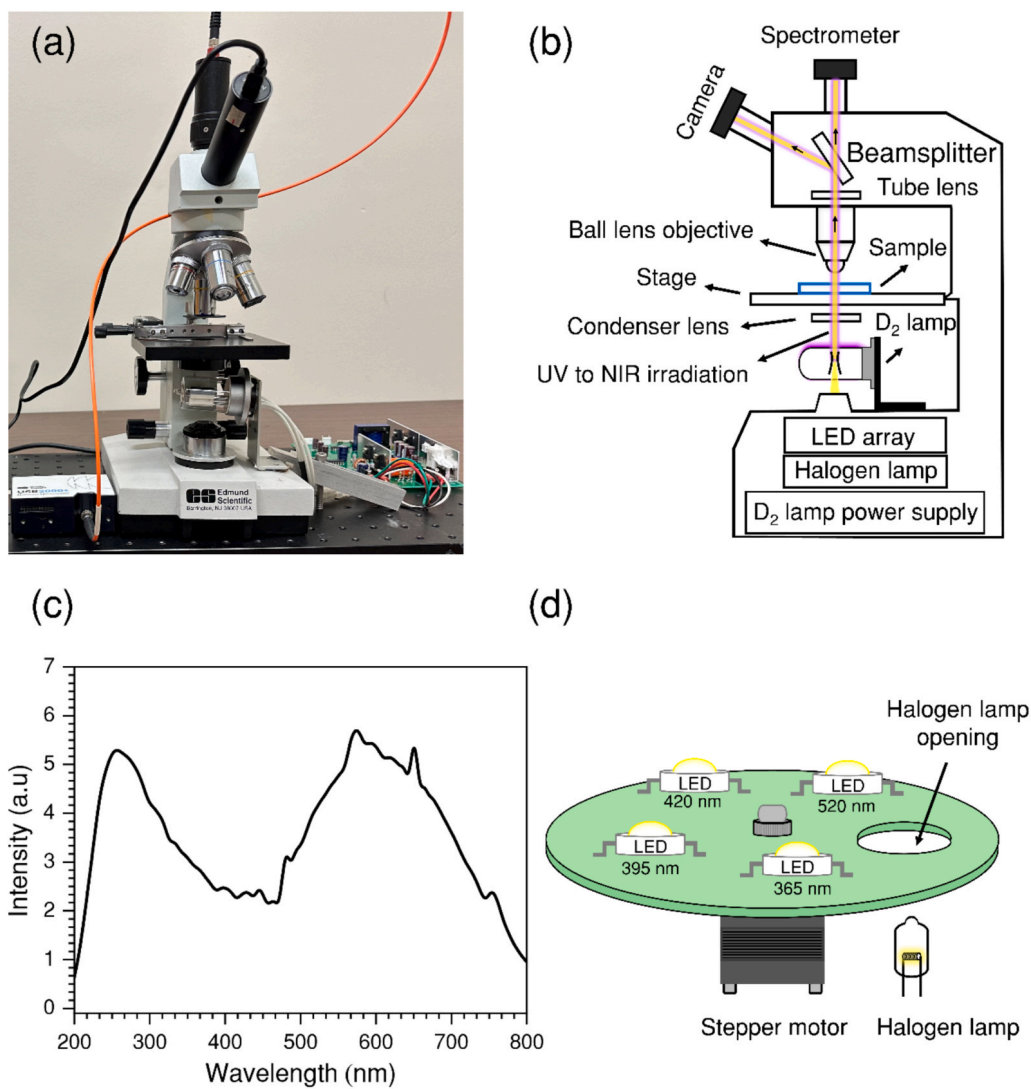

**Fig. 1.** UV–vis-NIR microspectrophotometer system; (a) Photograph of the constructed MSP setup; (b) Schematic diagram illustrating the key components and optical path of the system; (c) Continuous spectrum emitted by the $D_2$-halogen light source; (d) Detailed schematic of the LED array and the halogen lamp configuration.

*3.2. Comparative cost analysis*

A comprehensive cost analysis was conducted to compare the MSP system developed in this study with a commercially available microspectrophotometer offering similar capabilities. The selected commercial system integrates microscopy imaging with spectral measurements, supports both absorption and fluorescence measurements, and operates across a broad spectral range spanning the UV–vis-NIR regions, features that closely align with those of our MSP system. However, due to confidentiality agreements regarding the pricing information, the vendor's name and the specific model of the commercial system cannot be disclosed when providing a detailed cost breakdown.

Table 2 presents a cost breakdown of the core components of our MSP system compared to the selected commercial MSP, based on a quotation from the manufacturer. To ensure a transparent comparison, the analysis focuses on key design components that contributed most significantly to the compactness and cost-efficiency of our system. The spectrometer and camera were excluded from the comparison, as they are modular elements that can be independently replaced.

The prices listed for the commercial MSP in Table 2 represent those of a fully developed, commercially available product. For high-tech products, the manufacturing cost typically represents about one-third of the retail price [65]. Using this estimate, our calculation of a 20-fold reduction in construction cost is well-supported.

*3.3. Spatial resolution assessment*

Using the methods described in section 2.5, the fourth element of the sixth group on the USAF 1951 resolution target was identified as the smallest feature that could be clearly resolved by the system, as shown in Fig. 2(a). The corresponding intensity profile, demonstrating clear resolution of the line pattern, is presented in Fig. 2(b). Based on Eq. (1) and Eq. (2), this corresponds to a spatial resolution of approximately 11.04 μm in both horizontal and vertical directions.

With the MSP optical configuration, which includes a 4 mm half-ball objective lens, the resulting image FOV was measured to be approximately 220 μm × 180 μm, with magnification of ~ 43×.

The spectrally active region was determined to be approximately 70 μm × 70 μm, enabling micrometer-scale spatial resolution in selecting the area from which the spectra were acquired.

*3.4. Spectrophotometric and colorimetric tests*

To validate the spectrophotometric and colorimetric accuracy of this UV–vis-NIR MSP, an aqueous solution of Ru(bpy)$_3^{2+}$ was used as a test sample. Fig. 3 shows the normalized absorption spectrum of Ru(bpy)$_3^{2+}$ in aqueous solution, measured using both the Shimadzu UV-1601 spectrophotometer and the UV–vis–NIR MSP system. Prominent absorption bands are observed at 245 nm and 450 nm, along with a lower intensity band at 425 nm. These bands are attributed to metal-to-ligand charge transfer (MLCT) transitions involving the d orbitals of ruthenium (Ru) and the π* orbitals of bipyridine (bpy). The shoulder at 425 nm arises from the splitting of the first excited-state energy levels due to the complex's trigonal symmetry. Additionally, the lower-intensity absorption features at 325 nm and 350 nm correspond to metal-centered (MC) d-d transitions. The absorption bands at 200 nm and 285 nm are associated with ligand-centered (LC) π-π* transitions. These results closely match the data from the literature, confirming the accuracy of our MSP [66,67].

**Table 1**
List of components used in the MSP system, including their function and operating wavelength ranges.

| Subsystems | Function | Components | Operating wavelength range |
| --- | --- | --- | --- |
| Light sources | Absorption measurement | Deuterium lamp | 200-400 nm |
| | | Tungsten-halogen lamp | 400 nm-2.2 μm |
| | Fluorescence measurement | LEDs | 365, 395, 420, and 520 nm |
| Optics | Objective lens | Fused silica hemispherical lens | 200 nm-2 μm |
| | Beam splitter | UV fused silica plate | 185 nm-1.1 μm |
| | Condenser lens | Biconvex fused silica, fl = 35 mm | 185 nm-2.1 μm |
| | Tube lens | Biconvex fused silica, fl = 50 mm | 185 nm-2.1 μm |
| | Colored filter | Long-pass colored glass | 550 nm-1.8 μm |
| Imaging unit | Camera | CMOS | 400 nm-700 nm |
| | | UV-sensitive | 200 nm-400 nm |
| Spectral acquisition unit | Spectrometer | Spectrometer coupled to CCD | 200 nm-800 nm |
| Power supply | DC voltages: 24, 12, and 5 Volt | Rechargeable battery | Not applicable |

**Table 2**
Cost comparison between the proposed MSP and the commercial MSP system.

| System components | Estimated cost (Proposed MSP system) | Estimated cost (Commercial MSP system) |
| --- | --- | --- |
| Microscope body and optics | < $ 1000 | $ 65,000 |
| Light source for absorption measurements | $ 1000 | $ 27,000 |
| Light source for fluorescence measurements | < $ 100 | $ 29,000 |
| Microscope objective | < $ 50 | $ 7000 |
| Total cost | ~ $ 2150 | $ 128,000 |

Fig. 4 presents the colorimetric calibration curve, which shows the change in absorbance of Ru(bpy)$_3^{2+}$ aqueous solutions as a function of concentration, ranging from 10 μM to 1 mM. A and $A_0$ represent the absorbance of the aqueous solution of Ru(bpy)$_3^{2+}$ and distilled water at 450 nm, respectively. The absorbance measurements by the MSP exhibit a linear response to concentration changes, with an $R^2$ value of 0.998, demonstrating the high precision of the MSP in measuring absorbance. The maximum relative error observed was 6.86 %.

The LDR shown in Fig. 4 is influenced by several experimental parameters, including the optical properties of the analyte, solvent pH,

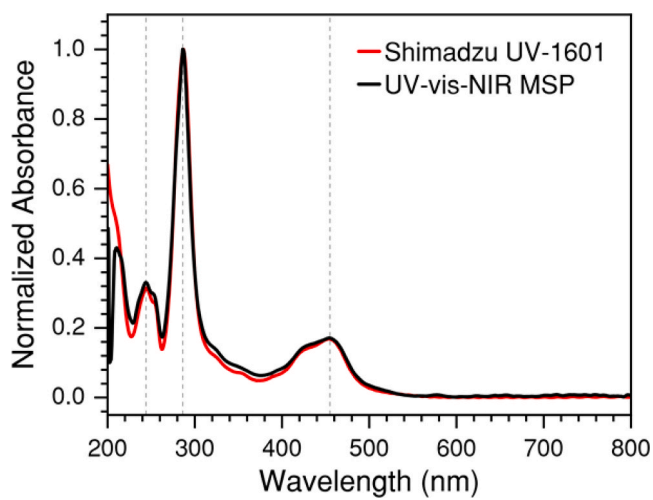

**Fig. 3.** Absorption spectra of Ru(bpy)$_3^{2+}$ aqueous solution recorded by a Shimadzu UV-1601 spectrophotometer and our recently developed UV–vis-NIR MSP.

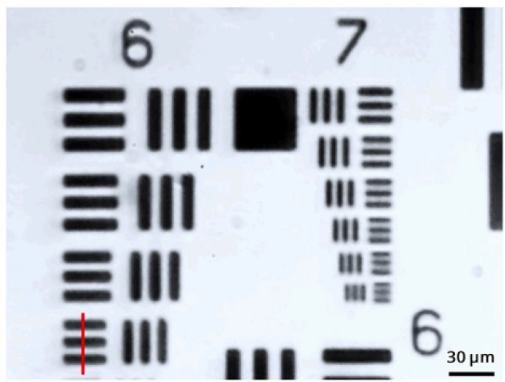
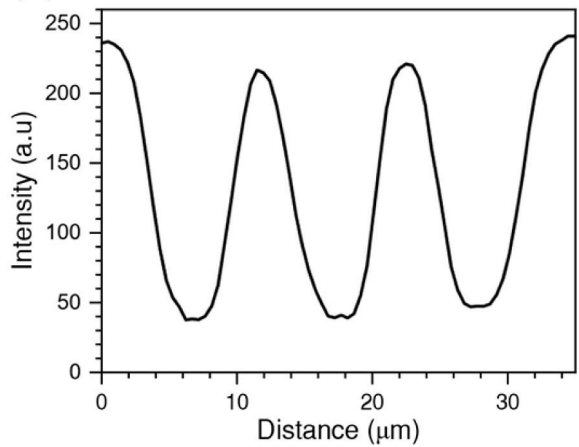

**Fig. 2.** (a) Microscope image of a USAF 1951 resolution target captured using our MSP system with a 4 mm diameter fused silica half-ball objective lens and a CMOS camera. (b) Line profile extracted along the red line at the fourth element of the sixth group. (For interpretation of the references to color in this figure legend, the reader is referred to the web version of this article.)

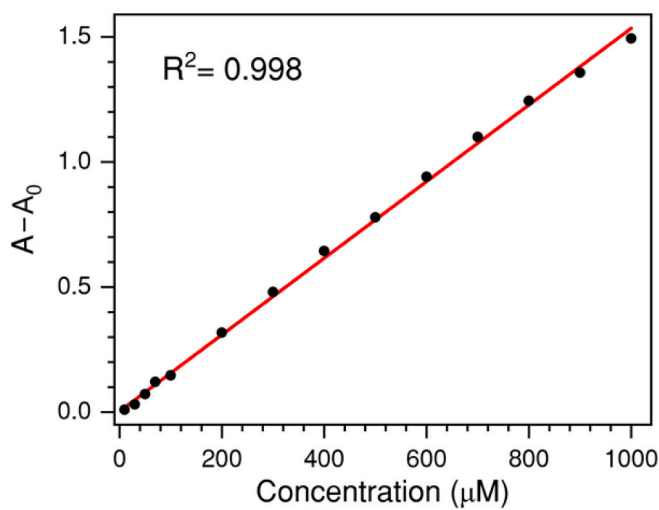

**Fig. 4.** Calibration plot of the absorbance at 450 nm versus the concentration of $Ru(bpy)_3^{2+}$ aqueous solution using the UV–vis-NIR MSP. A and $A_0$ represent the absorbance of the aqueous solution of $Ru(bpy)_3^{2+}$ and distilled water at 450 nm, respectively.

detection wavelength, cuvette path length, optical system configuration, and detector sensitivity. In our colorimetric measurements using the MSP setup with an aqueous solution of $Ru(bpy)_3^{2+}$ in a quartz cuvette with a 1 mm path length, the minimum detectable absorbance at 450 nm corresponded to a concentration of 10 μM. This detection limit could be further improved by using an analyte with a higher molar absorptivity, a cuvette with a longer optical path length, and a detector with enhanced sensitivity. The upper limit of the LDR was 1 mM. At higher concentrations, the optical density increased nonlinearly due to analyte aggregation. Adjusting the pH of the solvent may mitigate this effect and extend the linear range.

### 3.5. Spectrofluorometric and fluorometric tests

To evaluate the performance of our MSP in fluorescence measurements, an aqueous solution of $Ru(bpy)_3^{2+}$ was used as a test sample. Fig. 5 shows the normalized fluorescence spectra of $Ru(bpy)_3^{2+}$ aqueous solution excited at 420 nm, measured using both the Shimadzu RF-6000 spectrofluorometer and the UV–vis-NIR MSP. Fluorescence spectra show a maximum at approximately 620 nm, which is attributed to the transition from the triplet metal-to-ligand charge transfer ($^3$MLCT) excited state to the ground state [68]. Fig. 6 presents the fluorometric calibration curve, showing the fluorescence intensity as a function of $Ru(bpy)_3^{2+}$ concentrations ranging from 10 μM to 1 mM. F and $F_0$ represent the fluorescence intensity of the $Ru(bpy)_3^{2+}$ solution and distilled water, respectively. The fluorescence calibration curve was plotted on a log–log scale using the procedure described in [69]. The fluorescence intensity measured by the MSP exhibits a linear response to concentration changes, with an $R^2$ value of 0.990, which indicates the high accuracy of the instrument. The maximum relative error observed was 8.58 %.

In fluorometric measurements using the MSP setup with an aqueous solution of $Ru(bpy)_3^{2+}$ in a quartz cuvette with a 1 mm path length, the minimum detectable fluorescence was observed at a concentration of 10 μM. This detection limit could be improved by using a fluorophore with stronger fluorescence, employing a cuvette with a longer optical path length, and utilizing a more sensitive detector. The upper limit of the LDR was approximately 1 mM. At concentrations beyond this, the fluorescence intensity decreased due to self-quenching of the fluorophore. Modifying the pH of the solvent may reduce self-quenching and extend the linear range.

### 3.6. Absorption and fluorescence analysis of solid samples

The ability of our MSP to record absorption and fluorescence spectra from solid samples is demonstrated by the spectra acquired from the $Ru(bpy)_3^{2+}$ dye and bacteriorhodopsin proteins in crystalline form.

**$Ru(bpy)_3^{2+}$**: Fig. 7(a) and Fig. 7(b) present microscopic images of the dried form and a single crystal of $Ru(bpy)_3^{2+}$, respectively, prepared as previously described. The black box within the image represents the entrance aperture of the spectrometer, indicating the area from which the spectra were recorded.

UV–vis-NIR absorption spectra and fluorescence spectra of the crystal and dried forms of $Ru(bpy)_3^{2+}$ are compared with those of the aqueous solution, as shown in Fig. 7(c) and Fig. 7(d), respectively.

Both solid-state samples exhibit a bathochromic shift in their absorption band maxima compared to the solution, as illustrated in Fig. 7 (c). In the solid samples, the maximum MLCT and LC electronic transitions occur at 470 nm and 300 nm, respectively. The absorption peak at 245 nm in solution exhibits a minor redshift of around 5 nm in the crystal form, which increases to approximately 15 nm in the dried form. The MLCT absorption band at 260 nm in the dried sample shows higher intensity compared to the crystal MLCT absorption band at 250 nm. Fig. 7(c) highlights a distinct difference in the 450 nm MLCT and 285 nm LC ratios between the solution and solid states. The MLCT-to-LC ratio increases from 0.2 in solution to more than 0.7 in solid state, indicating a considerable increase in the absorption intensity of the MLCT electronic transitions relative to those of LC. The bathochromic shift in the absorption spectra is attributed to the increase in molecular density and aggregation in the solid state, which enhances intermolecular interactions and may alter the energies of the electronic transitions, thereby affecting the positions of the absorption maxima [62]. As shown in Fig. 7(d), the fluorescence spectra of the dried and crystalline samples show emission maxima at approximately 575 nm and 595 nm, respectively, both exhibiting a blue shift relative to the solution spectrum. This blue shift of the emission maxima is attributed to rigidochromism, which refers to an increase in the energy of the MLCT emission in solid media compared to solution media [70]. Upon excitation, the excited state of $Ru(bpy)_3^{2+}$ shows an asymmetric charge distribution, while the charge distribution in its ground state remains symmetric. In solution, the solvent molecules can reorient their dipoles to stabilize the polar excited state, facilitating thermal relaxation of the Franck–Condon excited state. Consequently, the complex emits light from a relaxed, lower energy state. In contrast, within a solid matrix, where solvent reorientation is limited, the excited state is only partially stabilized or relaxed during its lifetime, leading to emission from a higher energy state compared to that in solution. This results in a blue shift in the fluorescence emission spectrum. Interestingly, the fluorescence spectra of $Ru(bpy)_3^{2+}$ crystals exhibit a red shift compared to those of the dried form. The observed red shift aligns with previous findings for $Ru(bpy)_3^{2+}$ in solidified gel [71]. However, our experimental approach, which involved recording fluorescence spectra from single crystals of $Ru(bpy)_3^{2+}$ without any additional materials incorporated, suggests an alternative interpretation. Our results indicate that the mechanism responsible for this red shift is inherent to the $Ru(bpy)_3^{2+}$ complex, rather than due to solidification conditions, interactions with oxygen atoms from the gel network, or the porosity of the gel, as previously proposed [71].

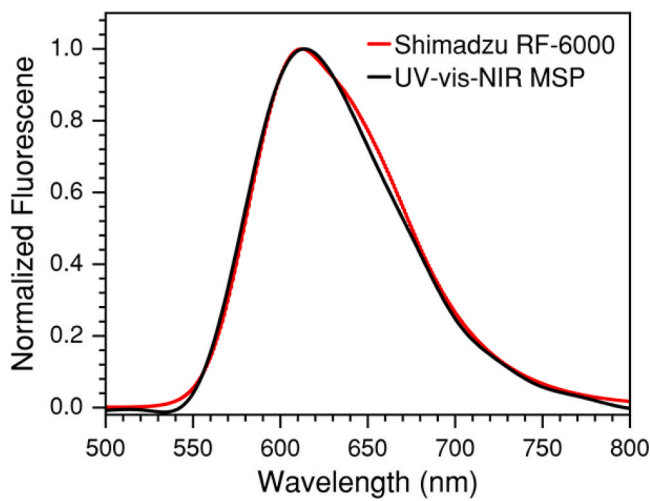

**Fig. 5.** Fluorescence spectra of the $Ru(bpy)_3^{2+}$ solution at an excitation wavelength of 420 nm, measured using both the Shimadzu RF-6000 spectrofluorometer and the UV–vis-NIR MSP.

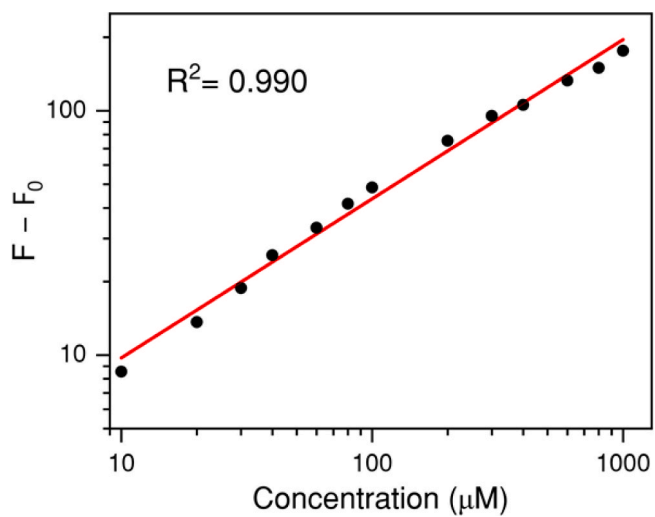

**Fig. 6.** Calibration plot of fluorescence intensity versus the concentration of $Ru(bpy)_3^{2+}$ using the presented UV–vis-NIR MSP. F and $F_0$ represent the fluorescence intensity of the $Ru(bpy)_3^{2+}$ solution and distilled water, respectively.

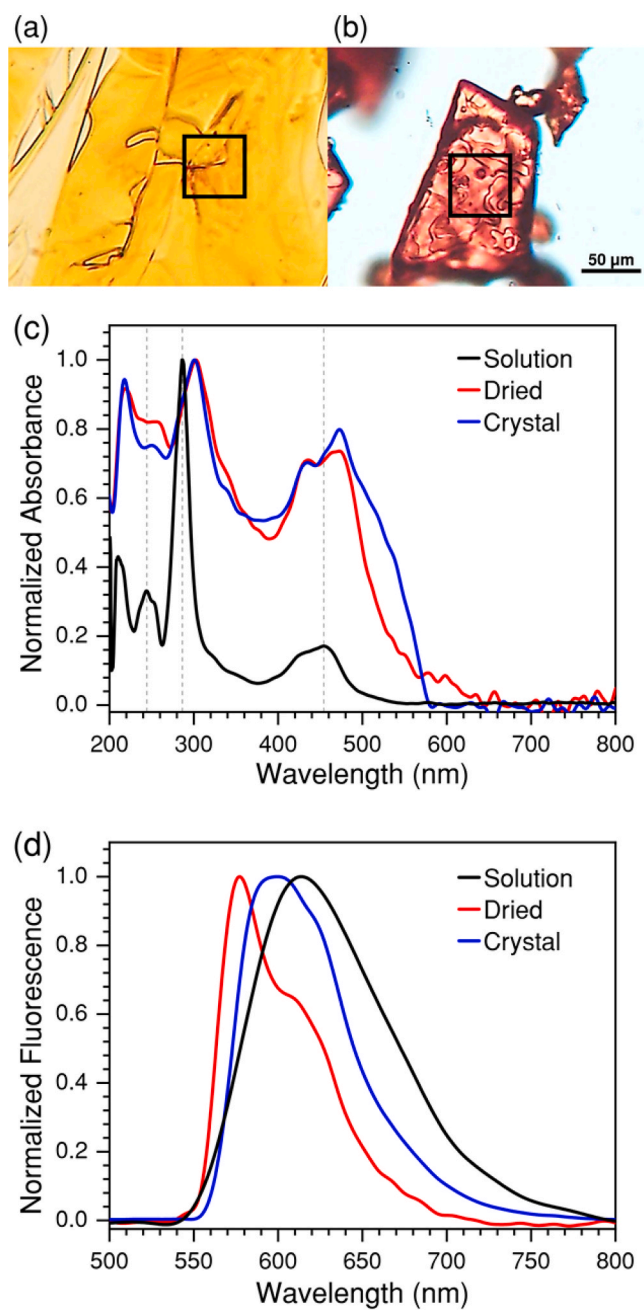

**Fig. 7.** Microscopic image of (a) dried $Ru(bpy)_3^{2+}$ (b) $Ru(bpy)_3^{2+}$ crystal; (c) absorption spectra and (d) fluorescence spectra of $Ru(bpy)_3^{2+}$ in solution, dried, and crystal forms, recorded using the UV–vis-NIR MSP.

**Bacteriorhodopsin:** Fig. 8(a) presents a microscopic image of bacteriorhodopsin crystals. The black box within the image represents the entrance aperture of the spectrometer, indicating the area from which the spectra were recorded. Fig. 8(b) presents the steady-state absorption spectra of the bacteriorhodopsin crystals, compared to those of the detergent buffer solution. The retinal chromophore absorption band, centered at 568 nm in detergent buffer solution, exhibits a slight blue shift of approximately 8 nm when measured in the crystalline form. Furthermore, bacteriorhodopsin crystals show a more pronounced absorption band near 400 nm compared to the solution, aligning with literature results [72]. The protein absorption band at 280 nm remains comparable between the solution and crystalline states, whereas the 220 nm band in the solution undergoes a red shift of about 10 nm in the crystalline form. The absorption band ratio at 560 nm to 280 nm increases significantly from approximately 0.3 in solution to 0.9 in the crystalline form, indicating increased absorption of retinal chromophore in the solid state.

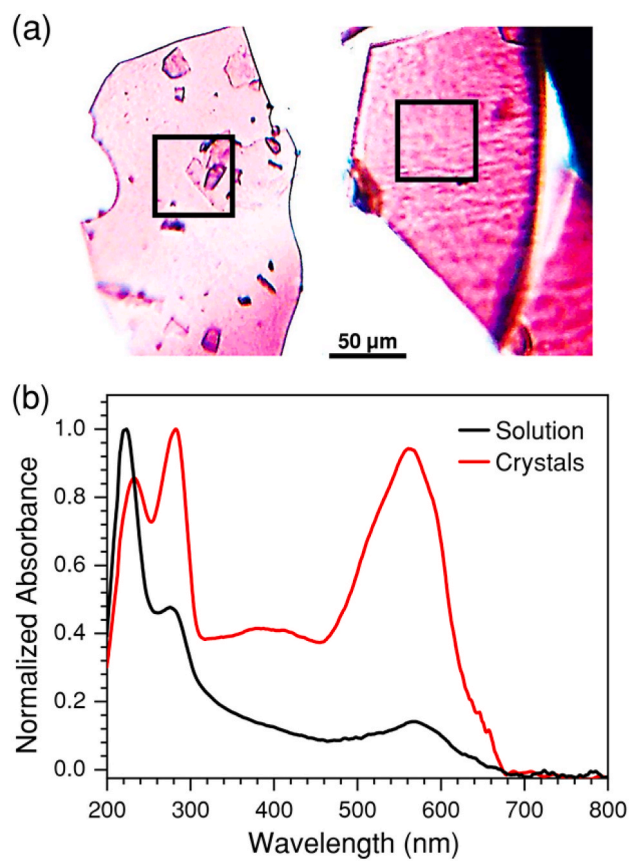

**Fig. 8.** (a) Microscopic image of bacteriorhodopsin crystals; (b) absorption spectra of bacteriorhodopsin crystals and detergent buffer solution recorded by UV–vis-NIR MSP.

### 3.7. Absorption and fluorescence analysis of living cells

To extend application of our MSP to biological samples and evaluate its suitability for in vivo experiments, absorption and fluorescence spectra of a single living cell of C. reinhardtii were recorded. Fig. 9(a) presents a microscopic image of a single living C. reinhardtii cell captured using a 100× objective lens. Fig. 9(b) and Fig. 9(c) show the UV–vis-NIR absorption and fluorescence spectra, respectively, recorded from the entire single cell using a 4 mm diameter fused silica hemispherical objective lens. The absorption spectra show prominent bands at approximately 430 nm and 670 nm. The band at 430 nm results from the presence of carotenoids, chlorophyll $a$, and chlorophyll $b$. The band at 670 nm is attributed to chlorophylls $a$ and $b$ [73,74]. The fluorescence spectra of the living C. reinhardtii cell show a peak at about 690 nm which is mostly attributed to chlorophyll $a$.

### 3.8. Assessment of photodamage risk

The interaction of optical radiation with samples during spectroscopic and microscopic analyses can induce a range of photodamage effects, including photochemical, photothermal, and photooxidative alterations, which may compromise sample integrity and the reliability of experimental results. These effects are particularly pronounced in sensitive biological, dried, or crystalline samples, where even moderate illumination may cause irreversible changes. Consequently, evaluating and mitigating photodamage is essential for obtaining reliable spectroscopic and microscopic data. Such effects are influenced by multiple factors, including the wavelength, intensity, and duration of illumination, as well as the optical and thermal properties of the sample itself. Understanding these parameters is crucial for minimizing damage while maintaining sufficient signal intensity for accurate analysis.

In our custom-built MSP, irradiance of each light source was measured at the sample plane under standard operating conditions. Specifically, the $D_2$ lamp delivered an irradiance of 0.84 mW/cm$^2$, the tungsten–halogen lamp emitted 1.57 mW/cm$^2$, and the combined output when both lamps were on provided 2.30 mW/cm$^2$. For fluorescence excitation, the LEDs provided irradiances of 12.1 mW/cm$^2$ (365 nm), 4.9 mW/cm$^2$ (395 nm), 8.8 mW/cm$^2$ (420 nm), and 4.7 mW/cm$^2$ (520 nm). High-power LEDs were deliberately selected as fluorescence excitation sources to provide high irradiance. This is necessary because fluorescence is inherently a low-efficiency process, only a small fraction of absorbed photons are re-emitted as detectable fluorescence. Therefore, higher excitation intensity is required to generate sufficient fluorescence signal above the detection threshold, particularly for weakly fluorescing samples. Subsequently, our measured irradiance values fall within the practical illumination levels commonly used in absorption and fluorescence spectroscopy [75–78].

To further ensure that photodamage did not occur during our measurements, a series of control experiments were conducted. Specifically, for each light source, spectra were recorded from the exact same spot on the solid Ru(bpy)$_3^{2+}$ samples before and after five minutes of continuous illumination. The results of these control experiments showed no sig-nificant change in the spectra, indicating that no detectable photodamage occurred due to prolonged illumination. This observation demonstrates that the irradiance levels employed in our setup were safe for the samples under the experimental conditions. Nevertheless, the extent of photodamage is inherently sample-dependent, determined by the optical absorption properties, thermal conductivity, and chemical stability of the material. Transparent samples with low absorbance may remain unaffected even under high irradiance, while strongly absorbing materials, such as dyes, proteins, or crystalline compounds, can undergo rapid degradation or structural changes. Dried and crystalline samples, in particular, are more susceptible to photothermal effects due to their limited heat dissipation capacity, whereas biological samples are especially prone to photochemical and photooxidative damage, particularly under UV illumination. Therefore, it is essential to use carefully controlled illumination strategies that are optimized for each sample type to ensure sample integrity and maintain the accuracy of spectroscopic measurements. To minimize the risk of photodamage, particularly for sensitive samples, several mitigation strategies can be considered. One approach involves carefully adjusting the irradiance of each light source by introducing neutral density (ND) filters into the illumination path to maintain the lowest practical intensity that still provides sufficient signal strength. Additionally, minimizing exposure durations can help to prevent prolonged light exposure, further reducing the cumulative energy delivered to the samples. Beam spreading is another effective method for lowering local irradiance. Specifically, the condenser lens in the optical pathway provides a more uniform and diffuse illumination over the sample surface, preventing localized hotspots that may cause photothermal damage. These combined strategies help ensure that samples are exposed to the minimum necessary light dose, thereby preserving their structural integrity while maintaining reliable spectroscopic measurements.

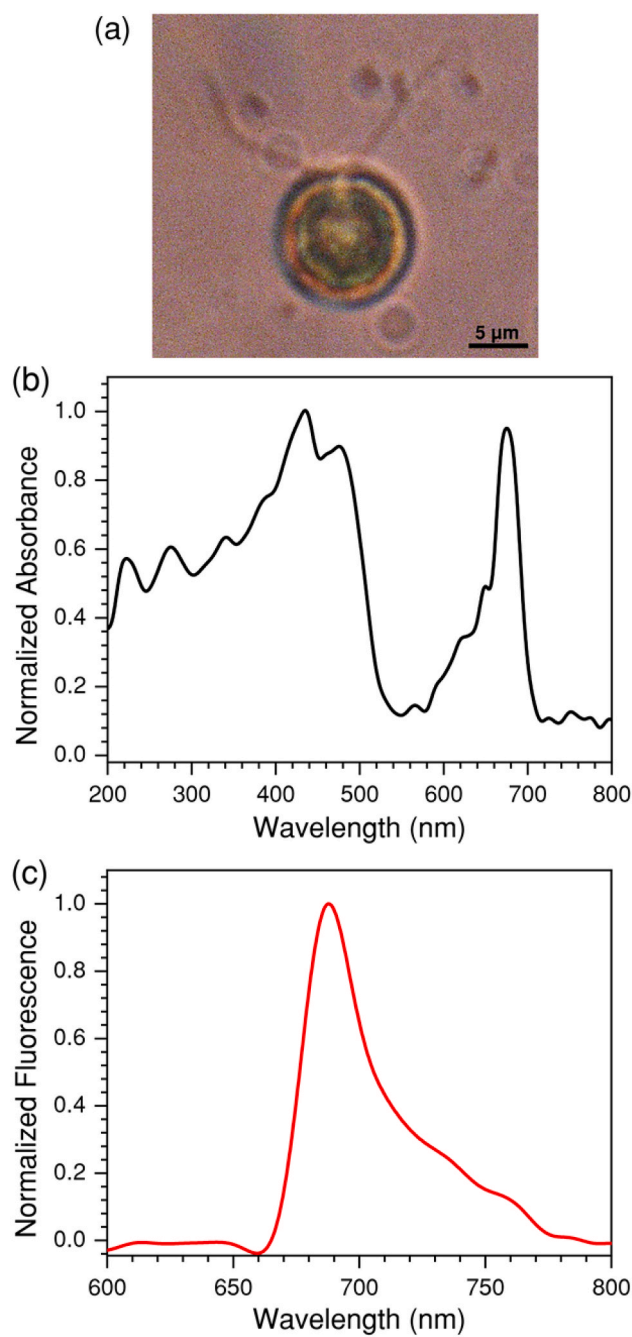

Fig. 9. Chlamydomonas reinhardtii single living cell; (a) microscopic image; (b) absorption spectrum; (c) fluorescence spectrum recorded using the UV–vis-NIR MSP.

## 4. Conclusion

In conclusion, a novel, portable, compact, cost-effective, and reliable microspectrophotometer has been developed, operating in the broad spectral range from 200 nm to 800 nm. This UV–vis-NIR MSP is capable of recording high-resolution absorption and fluorescence spectra in a single compact system. The dual-head design of this system enables the simultaneous monitoring of micron-sized heterogeneous samples and the recording of their spectra in situ and in real time. Our MSP design not only demonstrates compactness and portability but also achieves a 20-fold reduction in cost relative to comparable benchtop commercial microscope-spectrophotometer systems. The spectrophotometry and spectrofluorometric performance of our MSP was validated through comparative studies with commercial systems. The R-squared values of 0.998 and 0.990 were obtained by colorimetric and fluorescent tests, respectively, indicating the high accuracy of the MSP in measuring absorbance and fluorescence intensity. In addition, the applicability of our system has been demonstrated by recording the absorption and fluorescence spectra of a wide variety of samples, including dyes, protein crystals in both liquid and solid forms, and living cells. By addressing the limitations of conventional MSPs, such as high cost, large system size, and complexity, our portable MSP is particularly useful for rapid, on-site applications in various fields, including pathogen detection (bacteria and fungi), forensic analysis, material characterization, environmental pollution monitoring, cancer research, DNA analysis, microfluidic analysis, protein crystal studies, and applications in the semiconductor, food, and pharmaceutical industries.

## CRediT authorship contribution statement

**Negar Karpourazar:** Writing – review & editing, Writing – original draft, Visualization, Validation, Methodology, Investigation, Formal analysis, Data curation, Conceptualization. **Keyvan Khosh Abady:** Writing – review & editing. **Peter M. Rentzepis:** Writing – review & editing, Supervision, Resources, Project administration, Funding acquisition, Conceptualization.

## Declaration of competing interest

The authors declare the following financial interests/personal relationships which may be considered as potential competing interests: Peter M. Rentzepis reports financial support was provided by Air Force Office of Scientific Research. Peter M. Rentzepis, Negar Karpourazar, Keyvan Khosh Abady has patent pending to Texas A&M University System. If there are other authors, they declare that they have no known competing financial interests or personal relationships that could have appeared to influence the work reported in this paper.

## Acknowledgments

This study was supported by Air Force Office of Scientific Research (AFOSR) Grant number FA9550-20-1-0139 and Texas A&M Engineering Experiment Station (TEES) funds.

## Data availability

Data will be made available on request.

## References

[1] M. Karabaliev, B. Tacheva, B. Paarvanova, R. Georgieva, Change in osmotic pressure influences the absorption spectrum of hemoglobin inside red blood cells, Cells 13 (7) (2024), https://doi.org/10.3390/cells13070589.

[2] M.L. Bols, et al., In situ UV-Vis-NIR absorption spectroscopy and catalysis, Chem. Rev. 124 (5) (2024) 2352–2418, https://doi.org/10.1021/acs.chemrev.3c00602.

[3] C. Erkmen, I. Celik, Interaction mechanism of a pesticide, Azoxystrobin with bovine serum albumin: Assessments through fluorescence, UV-Vis absorption, electrochemical and molecular docking simulation techniques, Spectrochim. Acta A Mol. Biomol. Spectrosc., vol. 308, p. 123719, Mar 5 2024, doi: 10.1016/j.saa.2023.123719.

[4] L. Chen, et al., A review: Comprehensive investigation on bandgap engineering under high pressure utilizing microscopic UV–Vis absorption spectroscopy, APL Mater. 12 (3) (2024), https://doi.org/10.1063/5.0196152.

[5] K.K. Abady, N. Karpourazar, A. Krishnamoorthi, R. Li, P.M. Rentzepis, Spectroscopic analysis of bacterial photoreactivation, Photochem. Photobiol., Aug 29 2024, doi: 10.1111/php.14019.

[6] A. Krishnamoorthi, K. Khosh Abady, D. Dhankhar, P. M. Rentzepis, "Ultrafast Transient Absorption Spectra and Kinetics of Rod and Cone Visual Pigments," Molecules, 28(15), Aug 2 2023, doi: 10.3390/molecules28155829.

[7] G. Pompidor, F.S. Dworkowski, V. Thominet, C. Schulze-Briese, M.R. Fuchs, A new on-axis micro-spectrophotometer for combining Raman, fluorescence and UV/Vis absorption spectroscopy with macromolecular crystallography at the Swiss Light Source, J. Synchrotron Radiat. 20 (Pt 5) (Sep 2013) 765–776, https://doi.org/10.1107/S0909049513016063.

[8] L.M. Miller, P. Dumas, From structure to cellular mechanism with infrared microspectroscopy, Curr. Opin. Struct. Biol. 20 (5) (Oct 2010) 649–656, https://doi.org/10.1016/j.sbi.2010.07.007.

[9] A.N. Rousseau, et al., Fast antibiotic susceptibility testing via Raman microspectrometry on single bacteria: an MRSA case study, ACS Omega 6 (25) (2021) 22–26, https://doi.org/10.1021/acsomega.1c00170.

[10] K.S. Lee, Z. Landry, F.C. Pereira, M. Wagner, D. Berry, W.E. Huang, R. Stocker, Raman microspectroscopy for microbiology, Nat. Rev. Methods Primers (2021).

[11] B. Sorli, et al., Micro-spectrometer for NMR: analysis of small quantities in vitro, Meas. Sci. Technol. 15 (5) (2004) 877–880, https://doi.org/10.1088/0957-0233/15/5/013.

[12] D. Bourgeois, X. Vernede, V. Adam, E. Fioravanti, T. Ursby, A microspectrophotometer for UV–visible absorption and fluorescence studies of protein crystals, J. Appl. Cryst. 35 (3) (2002) 319–326.

[13] C. Hu, H. Mei, H. Guo, J. Zhu, Color analysis of textile fibers by microspectrophotometry, Forensic Chem. 18 (2020), https://doi.org/10.1016/j.forc.2020.100221.

[14] P. Martin, Spectroscopy with a light optical microscope, Microscopy Today 21 (1) (2012) 22–26, https://doi.org/10.1017/s1551929512001034.

[15] P.C. Martin, M.B. Eyring, *Microspectrophotometry*. Experimental Methods in the Physical Sciences, Academic Press, 2014.

[16] Y. Zhu, J.X. Cheng, Transient absorption microscopy: Technological innovations and applications in materials science and life science, p. 020901, Jan 14, J. Chem. Phys. 152 (2) (2020), https://doi.org/10.1063/1.5129123.

[17] L. Barsanti, L. Birindelli, F. Sbrana, G. Lombardi, P. Gualtieri, Advanced microscopy techniques for molecular biophysics, Int. J. Mol. Sci. 24 (12) (2023), https://doi.org/10.3390/ijms24129973. Jun 9 2023.

[18] L. Barsanti, V. Evangelista, V. Passarelli, A.M. Frassanito, P. Gualtieri, Fundamental questions and concepts about photoreception and the case of Euglena gracilis, Integr. Biol. (Camb) 4 (1) (Jan 2012) 22–36, https://doi.org/10.1039/c1ib00115a.

[19] G.K.S.J.J. Wolken, Microspectrophotometry of euglena chloroplast and eyespot, Nature (1960).

[20] A.L. Forster, J.L. Bitter, S. Rosenthal, S. Brooks, S.S. Watson, Photofading in cotton fibers dyed using red, yellow, and blue direct dyes during examination with microspectrophotometry (MSP), Forensic Chem. 5 (Sep 2017) 72–78, https://doi.org/10.1016/j.forc.2017.06.006.

[21] D. Smigiel-Kaminska, J. Was-Gubala, P. Stepnowski, J. Kumirska, The identification of cotton fibers dyed with reactive dyes for forensic purposes, Molecules 25 (22) (2020), https://doi.org/10.3390/molecules25225435. Nov 20.

[22] K. Brinsko-Beckert, S. Palenik, O.R. Abraham, E. Groves, C.S. Palenik, Microscopical recognition and characterization of solution dyed fibers, J. Forensic Sci. 69 (1) (Jan 2024) 60–80, https://doi.org/10.1111/1556-4029.15423.

[23] A.R. Pearson, A. Mozzarelli, G.L. Rossi, Microspectrophotometry for structural enzymology, Curr. Opin. Struct. Biol. 14 (6) (Dec 2004) 656–662, https://doi.org/10.1016/j.sbi.2004.10.007.

[24] E. Gomez-Otero, M. Costas, I. Lavilla, C. Bendicho, Ultrasensitive, simple and solvent-free micro-assay for determining sulphite preservatives (E220-228) in foods by HS-SDME and UV-vis micro-spectrophotometry, Anal. Bioanal. Chem. 406 (8) (Mar 2014) 2133–2140, https://doi.org/10.1007/s00216-013-7293-3.
[25] C. E. M. Haishan Zeng, David I. McLean M.D., Branko Palcic, "Novel microspectrophotometer and its biomedical applications," Opt. Eng., vol. 32, 1993.
[26] E.S.P.L. Yan Jun Liu*, Jing Hua Teng, Dispersion-flattened transmission based on liquidcrystal-coated plasmonic subwavelength structures, Photonics Global Conference (2012).
[27] A.E. Nikolaenko, et al., Carbon nanotubes in a photonic metamaterial, Phys. Rev. Lett. 104 (15) (2010), https://doi.org/10.1103/PhysRevLett.104.153902, 153902, Apr 16.
[28] A. Gaspar, I. Bacsi, E.F. Garcia, M. Braun, F.A. Gomez, Application of external micro-spectrophotometric detection to improve sensitivity on microchips, Anal. Bioanal. Chem. 395 (2) (Sep 2009) 473–478, https://doi.org/10.1007/s00216-009-2980-9.
[29] K.I. Inoue, J. Mao, R. Okamoto, Y. Shibata, W. Song, S. Ye, Development of line-detected UV-Vis absorption microscope and its application to quantitative evaluation of lithium surface reactivity, Anal. Chem. 95 (9) (2023) 4550–4555, https://doi.org/10.1021/acs.analchem.2c05759.
[30] G.K.S.A.J.J. Wolken, "<science.130.3382.1084.pdf>," Science, 1959.
[31] J.A. Kim, D.J. Wales, G.-Z. Yang, Optical spectroscopy for in vivo medical diagnosis—a review of the state of the art and future perspectives, Prog. Biomed. Eng. 2 (4) (2020), https://doi.org/10.1088/2516-1091/abaaa3.
[32] J. Sorensen, et al., Online fluorescence spectroscopy for the real-time evaluation of the microbial quality of drinking water, Water Res. 137 (2018) 301–309.
[33] C. Vogt, C.S. Wondergem, B.M. Weckhuysen, Ultraviolet-Visible (UV-Vis) Spectroscopy, in: I.E. Wachs, M.A. Bañares (Eds.), Springer Handbook of Advanced Catalyst Characterization, Springer International Publishing, Cham, 2023, pp. 237–264.
[34] Z. Yu, R. Meng, S. Deng, L. Jia, An open-source handheld spectrometer for colorimetric and fluorescence analyses, Spectrochim. Acta A Mol. Biomol. Spectrosc. 287 (2) (2023), https://doi.org/10.1016/j.saa.2022.122072, p. 122072, Feb 15.
[35] X. Fang, W. Zhang, H. Su, W. Xie, L. Jia, A LAMP-based colorimetric and fluorescence dual-channel assay for on-site identification of adulterated meat by a portable device, Food Control 169 (2025) 110984.
[36] D. Dhankhar, P.M. Rentzepis, Techniques for constructing ultra low cost deep ultraviolet transmission microscopes, IEEE Access 11 (2023) 41964–41969, https://doi.org/10.1109/ACCESS.2023.3270570.
[37] L.S. Ferraraccio, P. Bertoncello, Electrochemiluminescence (ECL) biosensor based on tris(2,2'-bipyridyl)ruthenium(II) with glucose and lactate dehydrogenases encapsulated within alginate hydrogels, Bioelectrochemistry 150 (Apr 2023) 108365, https://doi.org/10.1016/j.bioelechem.2023.108365.
[38] B. Huang, C. Yao, Y. Zhang, X. Lu, A novel label-free solid-state electrochemiluminescence sensor based on the resonance energy transfer from Ru (bpy)(3)(2+) to GO for DNA hybridization detection, Talanta 218 (2020), https://doi.org/10.1016/j.talanta.2020.121126, 121126, Oct 1.
[39] Y. Luo, Z.Y. Xu, H. Wang, X.W. Sun, Z.T. Li, D.W. Zhang, Porous Ru(bpy)(3)(2+)-linked polymers for recyclable photocatalysis of enantioselective alkylation of aldehydes, ACS Macro Lett. 9 (1) (2020) 90–95, https://doi.org/10.1021/acsmacrolett.9b00872.
[40] A.P. Perrino, A. Miyagi, S. Scheuring, Single molecule kinetics of bacteriorhodopsin by HS-AFM, 7225, Dec 10, Nat Commun 12 (1) (2021), https://doi.org/10.1038/s41467-021-27580-2.
[41] S.P. Balashov, Protonation reactions and their coupling in bacteriorhodopsin, Biochim. et Biophys. Acta (BBA)-Bioenergetics 1460 (1) (2000) 75–94.
[42] N. Hampp, Bacteriorhodopsin as a photochromic retinal protein for optical memories, Chem. Rev. 100 (5) (2000) 1755–1776.
[43] P. Singh, S. Singh, N. Jaggi, K.-H. Kim, P. Devi, Recent advances in bacteriorhodopsin-based energy harvesters and sensing devices, Nano Energy 79 (2021), https://doi.org/10.1016/j.nanoen.2020.105482.
[44] J.A. Stuart, D.L. Marcy, K.J. Wise, R.R. Birge, Biomolecular electronic device applications of bacteriorhodopsin: three-dimensional optical & holographic associative memories, integrated biosensors. Molecular Electronics, Bio-Sensors and Bio-Computers: Springer (2003) 265–299.
[45] C. Espinoza-Araya, et al., A bacteriorhodopsin-based biohybrid solar cell using carbon-based electrolyte and cathode components, Biochim. Biophys. Acta Bioenerg. 1864 (4) (2023), https://doi.org/10.1016/j.bbabio.2023.148985, p. 148985, Nov 1.
[46] T.J. Johnson, S. Gakhar, S.H. Risbud, M.L. Longo, Development and characterization of titanium dioxide gel with encapsulated bacteriorhodopsin for hydrogen production, Langmuir 34 (25) (2018) 7488–7496, https://doi.org/10.1021/acs.langmuir.8b01471.
[47] M. Ahmadi, J.T. Yeow, Fabrication and characterization of a radiation sensor based on bacteriorhodopsin, Biosens. Bioelectron. 26 (5) (2011) 2171–2176, https://doi.org/10.1016/j.bios.2010.09.029.
[48] H. Hasegawa, et al., Biomaterial-based biomimetic visual sensors: inkjet patterning of bacteriorhodopsin, ACS Appl. Mater. Interfaces 15 (38) (2023) 45137–45145, https://doi.org/10.1021/acsami.3c07540.
[49] M.A. Scaife, G. Nguyen, J. Rico, D. Lambert, K.E. Helliwell, A.G. Smith, Establishing Chlamydomonas reinhardtii as an industrial biotechnology host, Plant J. 82 (3) (May 2015) 532–546, https://doi.org/10.1111/tpj.12781.
[50] V. Calatrava, M. Tejada-Jimenez, E. Sanz-Luque, E. Fernandez, A. Galvan, and A. Llamas, "Chlamydomonas reinhardtii, a reference organism to study algal-microbial interactions: why can't they be friends?," Plants (Basel), vol. 12, no. 4, Feb 9 2023, doi: 10.3390/plants12040788.
[51] E.H. Harris, The Chlamydomonas Sourcebook: Introduction to Chlamydomonas and Its Laboratory Use:, Volume 1, Academic press, 2009.
[52] C.M. Bellido-Pedraza, et al., Chlamydomonas reinhardtii, an algal model in the nitrogen cycle, Plants 9 (7) (2020) 903.
[53] C. Griesbeck, I. Kobl, M. Heitzer, Chlamydomonas reinhardtii: a protein expression system for pharmaceutical and biotechnological proteins, Mol. Biotechnol. 34 (2006) 213–223.
[54] G. Torzillo, A. Scoma, C. Faraloni, L. Giannelli, Advances in the biotechnology of hydrogen production with the microalga Chlamydomonas reinhardtii, Crit. Rev. Biotechnol. 35 (4) (2015) 485–496.
[55] Y. Milrad, V. Nagy, T. Elman, M. Fadeeva, S.Z. Tóth, I. Yacoby, A PSII photosynthetic control is activated in anoxic cultures of green algae following illumination, Commun. Biol. 6 (1) (2023) 514.
[56] M. Soudi, et al., Self-assembled plasmonic structural color colorimetric sensor for smartphone-based point-of-care ammonia detection in water, ACS Appl. Mater. Interfaces 16 (34) (2024) 45632–45639.
[57] M. Soudi, P. Cencillo-Abad, D. Chanda, Scalable plasmonic colorimetric sensor for on-site detection, in Nanoscale Imaging, Sensing, and Actuation for Biomedical Applications XXII, 2025, vol. 13335: SPIE, pp. 17-22.
[58] S. Qian, Y. Cui, Z. Cai, L. Li, Applications of smartphone-based colorimetric biosensors, Biosens. Bioelectron.: X 11 (2022), https://doi.org/10.1016/j.biosx.2022.100173.
[59] T. Samanta, R. Shunmugam, Colorimetric and fluorometric probes for the optical detection of environmental Hg(ii) and As(iii) ions, Mater. Adv. 2 (1) (2021) 64–95, https://doi.org/10.1039/d0ma00521e.
[60] M. Shimizu, T. Hiyama, Organic fluorophores exhibiting highly efficient photoluminescence in the solid state, Chem. Asian J. 5 (7) (2010) 1516–1531, https://doi.org/10.1002/asia.200900727.
[61] P. Boden, et al., NIR-emissive chromium(0), molybdenum(0), and tungsten(0) complexes in the solid state at room temperature, Chemistry 27 (51) (2021) 12959–12964, https://doi.org/10.1002/chem.202102208.
[62] E.L. Sciuto, et al., Photo-physical characterization of fluorophore Ru(bpy) 3 2+ for optical biosensing applications, Sens. Bio-Sens. Res. 6 (2015) 67–71, https://doi.org/10.1016/j.sbsr.2015.09.003.
[63] J.R.J. Dominik Heger, †, Petr Klán*,†, Aggregation of methylene blue in frozen aqueous solutions studied by absorption spectroscopy, J. Phys. Chem. A (2005).
[64] R. Kurrey, et al., A KBr-impregnated paper substrate as a sample probe for the enhanced ATR-FTIR signal strength of anionic and non-ionic surfactants in an aqueous medium, RSC Adv. 10 (66) (2020) 40428–40441, https://doi.org/10.1039/d0ra07286a.
[65] K.T. Ulrich, S.D. Eppinger, Product design and development, McGraw-Hill, 2016.
[66] V.B.A. Juris, F. Barigelletti, S. Campagna, P. Belser, A. von Zelewsky, Ru(II) polypyridine complexes: photophysics, photochemistry, eletrochemistry, and chemiluminescence, Elsevier 84 (1988) 85–277.
[67] P. Dongare, B.D.B. Myron, L. Wang, D.W. Thompson, T.J. Meyer, [Ru(bpy)3]2+* revisited. is it localized or delocalized? how does it decay? Coord. Chem. Rev. 345 (2017) 86–107, https://doi.org/10.1016/j.ccr.2017.03.009.
[68] A.C. Bhasikuttan, M. Suzuki, S. Nakashima, T. Okada, Ultrafast fluorescence detection in Tris (2, 2 '-bipyridine) ruthenium (II) complex in solution: relaxation dynamics involving higher excited states, J. Am. Chem. Soc. 124 (28) (2002) 8398–8405.
[69] ASTM Designation: E 578 - 07: Standard Test Method for Linearity of Fluorescence Measuring Systems.
[70] J.S. McCarthy, et al., Role of the trifluoropropynyl ligand in blue-shifting charge-transfer states in emissive Pt diimine complexes and an investigation into the PMMA-imposed rigidoluminescence and rigidochromism, Inorg. Chem. 61 (29) (2022) 11366–11376.
[71] P. Innocenzi, H. Kozuka, T. Yoko, Fluorescence properties of the Ru (bpy) 32+ complex incorporated in sol– gel-derived silica coating films, J. Phys. Chem. B 101 (13) (1997) 2285–2291.
[72] R. Efremov, V. Gordeliy, J. Heberle, G. Büldt, Time-resolved microspectroscopy on a single crystal of bacteriorhodopsin reveals lattice-induced differences in the photocycle kinetics, Biophys. J . 91 (4) (2006) 1441–1451.
[73] R. Bauer, L. Szalay, E. Tombacz, Migration of electronic energy from chlorophyll b to chlorophyll a in solutions, Biophys. J. 12 (7) (1972) 731–745.
[74] M.C. Rodríguez, L. Barsanti, V. Passarelli, V. Evangelista, V. Conforti, P. Gualtieri, Effects of chromium on photosynthetic and photoreceptive apparatus of the alga Chlamydomonas reinhardtii, Environ. Res. 105 (2) (2007) 234–239.
[75] I. Thorlabs, "Collimated LED Light Sources for Microscopy: M470L4-C4," 2021. [Online]. Available: https://www.thorlabs.com/catalogpages/obsolete/2021/M470L4-C4.pdf.
[76] J. O. Sochi, et al., "Increased fluorescence observation intensity during the photodynamic diagnosis of deeply located tumors by fluorescence photoswitching of protoporphyrin IX," J. Biomed. Opt., vol. 28, no. 5, p. 055001, 5/1 2023, doi: 10.1117/1.JBO.28.5.055001.
[77] H. Kim, et al., "Doxorubicin-fucoidan-gold nanoparticles composite for dualchemo-photothermal treatment on eye tumors," Oncotarget; Vol 8, No 69, 2017. [Online]. Available: https://www.oncotarget.com/article/23092/text/.
[78] L. Yuanhua, X. Xiang, S. Fei, G. Zhiyong, J. Dayong, Contrast-enhanced fluorescence microscope by LED integrated excitation cubes, Light Adv. Manuf. 4 (2) (2023) 94–103, https://doi.org/10.37188/lam.2023.008.